\documentclass[preprint]{revtex4}

\input epsf
\usepackage{amsmath,amssymb}
\usepackage{graphicx}

\draft

\begin{document}
\titlepage
\title{Note on the Radion Effective Potential in the Presence of Branes}
\author{Peng Wang$^1$\footnote{E-mail: pewang@eyou.com} and Xin-He Meng$^{1,2}$ \footnote{E-mail: xhm@
physics.arizona.edu}} \affiliation{1. Department of Physics,
Nankai University, Tianjin, 300071, P.R.China \\ 2. Department of
Physics, University of Arizona, Tucson, AZ 85721}

\begin{abstract}
In String Theory compactification, branes are often invoked to get
the desired form of the radion effective potential. Current
popular way of doing this assumes that the introduction of branes
will not modify the background geometry in an important way. In
this paper, we show by an explicit example that at least in the
codimension 2 case, the gravitational backreaction of the brane
cannot be neglected in deriving the radion effective potential.
Actually, in this case, the presence of branes will have no effect
on the dynamics of radion.

\end{abstract}

\maketitle

Current quantum gravity theories, like String Theory, often invoke
extra dimensions as a basic ingredient of the nature of spacetime.
In order for those theories to be phenomenologically acceptable,
one commonly assumes that the extra dimensions are compactified
with a small and stabilized volume. So the study of
compactification mechanisms and their stability are of central
importance in those theories (see, e.g., Ref.\cite{silverstein}
for a recent review).

In this paper we will focus on the dynamics of the volume of the
extra dimensions. In many String Theory compactification scheme,
this will be the only moduli that is left unfixed (see, e.g.,
Ref.\cite{kklt}). For this we have to obtain the 4-dimensional
effective potential for the radion, which is the 4-dimensional
field that corresponds to dilatations of the compact dimensions.
Recently, the discussion of radion effective potential received a
lot of interest also because its central importance in exploring
the String Theory ``Landscape" \cite{susskind} and in the attempt
of constructing de Sitter vacuum in String Theory \cite{kklt} in
view of the recent cosmological observation that our Universe is
now accelerating \cite{obs}. Radion effective potential can also
help us construct models that may account for the current cosmic
acceleration such as the construction in Ref.\cite{odintsov}.

When discussing the radion stabilization problem, one possible way
to contribute a term to the radion effective potential is adding
branes (see, e.g., Refs.\cite{silverstein, giddings}). After
adding a $p$-brane, it is often assumed that the contribution to
the radion effective potential in the Einstein frame is of the
form
\begin{equation}
V_p(\phi)=\tilde{\sigma}\exp\left(-{2n+3-p\over n}\sqrt{{2n\over
n+2}}{\phi\over M_4}\right),\label{branepoten}
\end{equation}
where $\tilde{\sigma}$ is the 'renormalized' brane tension that
have absorbed the volume and other numerical factors, $n$ is the
number of extra dimensions, $M_4$ is the 4-dimensional reduced
Planck mass. The implicit assumption underlying such a form of
contribution is that the introduction of branes will not modify
the background geometry in an important way, i.e. the
gravitational backreaction of the branes can be neglected
\cite{giddings} (A recent important progress in String Theory just
adopted this contribution of branes to uplift the negative minimum
of the radion effective potential to a positive metastable minimum
\cite{kklt}).  In this paper, we will show that this assumption
cannot always be justified. At least when the spacetime dimension
is 6, the gravitational backreaction of the 3-branes on the
background geometry can \emph{never} be neglected. Actually, the
backreaction will just cancel the contribution (\ref{branepoten})
and thus branes will have no effect on the dynamics of radion. We
will prove our assertion by a well studied 6-dimensional example
\cite{carroll-sta} for which the exact solution in the presence of
branes is known \cite{carroll-foot}.

To begin with, following the common formalism when discussing
radion stabilization \cite{silverstein, giddings, carroll-sta},
let's decompose the coordinates into the four macroscopic
dimensions $x^\mu$ and the two extra dimensions $y^a$, and
consider an ansatz for which the geometry factorizes into a
4-dimensional metric $g_{\mu\nu}$ depending only on $x$ and a
2-dimensional metric $\gamma_{ab}$ depending only on $y$,
\begin{equation}
    ds^2 = G_{AB}\, dX^A dX^B = g_{\mu\nu}(x)dx^\mu dx^\nu
  +b^2(x)\gamma_{ab}(y) dy^a dy^b\ .
  \label{metric}
\end{equation}
where $A,B=0,...,5$, $\mu,\nu=0,...,3$ and $a,b=4,5$, $b$ is the
radion field.

The model we will discuss is a $\mathcal{S}^2$ compactification
manifold stabilized by a bulk cosmological constant and magnetic
flux. This model is a special example of the mechanism in
Ref.\cite{rubin} and has recently been discussed in detail in
Ref.\cite{carroll-sta, guenther}. The bulk lagrangian is
\begin{equation}
  S_6 = \int d^6X\sqrt{|G|}\, \left( {1\over 2}M_6^4 R - \lambda
  - {1\over 4} F_{AB} F^{AB}\right)\ ,
  \label{action}
\end{equation}
where $M_6$ is the 6-dimensional reduced Planck mass and $\lambda
$ is the 6-dimensional vacuum energy density. The 2-form field
strength takes the form $F_{ab}=\sqrt{|\gamma|}B_0\epsilon_{ab}$,
where $B_0$ is a constant and $\epsilon$ is the standard
antisymmetric tensor. Other components of $F_{AB}$ vanish
identically.

Let's define the Einstein frame metric $\tilde{g}_{\mu\nu}$ by the
conformal transformation
\begin{equation}
\tilde{g}_{\mu\nu}=b^2g_{\mu\nu},\label{conformal}
\end{equation}
and define a new field $\phi$ by
\begin{equation}
b=\exp(\phi/2M_4),\label{phi}
\end{equation}
where $M_4^2=M_6^4\mathcal{V}$ is the 4-dimensional reduced Planck
mass and $\mathcal{V}\equiv\int d^2y\sqrt{|\gamma|}$ is the volume
of the extra dimensions. It has been shown in
Ref.\cite{carroll-sta} that, after dimensional reduction, the
radion dynamics will be described in the Einstein frame by a
standard scalar field theory with the effective potential,
\begin{equation}
\tilde{V}=\lambda\mathcal{V}(-2e^{-2\phi/M_4}+e^{-\phi/M_4}+e^{-3\phi/M_4}).\label{poten}
\end{equation}
where we have set $\lambda={B_0^2\over2}$.

Now let's consider adding branes to the above scenario. In the
simplest case which is also the case considered in the current
literature \cite{silverstein, giddings, carroll-foot}, the branes
are described by Nambu-Goto action (see Ref.\cite{carter} for an
elegant review),
\begin{equation}
S_{NG}=-\sum_i\int d^4x\sqrt{|g|}\sigma_i\ ,\label{brane}
\end{equation}
where $i$ labels the branes, $\sigma_i$ and $X_i$ are the tension
and position of the $i$th brane, respectively. The energy-momentum
tensor of branes follow by varying $G^{AB}$ in the action
(\ref{brane})
\begin{equation}
  T_{AB}^b = -\sum_i{\sigma_i\over\sqrt{\gamma}}
  \left(\begin{matrix} g_{\mu\nu}  & 0 \cr 0 & 0 \end{matrix}\right)
  \delta^{(2)}(y_i)\ ,
\end{equation}

In Ref.\cite{carroll-foot}, Carroll and Guica considered two equal
tension branes located at the two poles of the $\mathcal{S}^2$
manifold. They showed that the brane tension will induce a deficit
angle in the extra dimension, thus $\gamma_{ab}$ will be of the
form
\begin{equation}
\gamma_{ab}dy^a dy^b=a_0^2(d\theta^2+\alpha^2\sin^2\theta
d\phi^2),\label{gamma}
\end{equation}
with $a_0$, the radius of the extra dimension, and $\alpha$ given
by
\begin{equation}
a_0^2=\frac{M_6^4}{2\lambda}\ ,\qquad
 \alpha = 1-{\sigma\over
2\pi M_6^4}.\label{solution}
\end{equation}

Then let's us consider the problem of radion stabilization in the
above scenario. From the ansatz (\ref{metric}) we have
\begin{equation}
\sqrt{|G|}=\sqrt{|g|}\sqrt{|\gamma|}b^2,\label{1}
\end{equation}
\begin{equation}
R[G]_{\mu\nu}=R[g]_{\mu\nu}-2b^{-1}\nabla_\mu\nabla_\nu
b,\label{R[G]1}
\end{equation}
\begin{equation}
R[G]_{ab}=R[\gamma]_{ab}-\gamma_{ab}[b\nabla^2b+(\nabla
b)^2],\label{R[G]2}
\end{equation}
\begin{equation}
R[G]=R[g]+b^{-2}R[\gamma]-4b^{-1}\nabla^2b-2b^{-2}(\nabla
b)^2.\label{2}
\end{equation}

The bulk energy-momentum tensor contains contributions from the
bulk cosmological constant and the gauge field,
\begin{equation}
  T^{bulk}_{AB} = T_{AB}^{\lambda} + T_{AB}^{F}\ ,\label{1.4}
\end{equation}
for which the explicit forms are
\begin{eqnarray}
  T_{AB}^{\lambda} &=& -\lambda\left(\begin{matrix} g_{\mu\nu} & 0 \cr
  0 & b^2\gamma_{ab} \end{matrix}\right)\cr
  T_{AB}^{F} &=& -{b^{-4}\over 2} B_0^2\left(\begin{matrix} g_{\mu\nu} & 0 \cr
  0 & -b^2\gamma_{ab} \end{matrix} \right)\ .\label{1.3}
\end{eqnarray}

Then, the longitudinal component of the Einstein equations is
\begin{eqnarray}
R[g]_{\mu\nu}-2b^{-1}\nabla_\mu\nabla_\nu b-{1\over
2}g_{\mu\nu}(R[g]+b^{-2}R[\gamma]-4b^{-1}\nabla^2 b-2b^{-2}(\nabla
b)^2)\cr={1\over M_6^4}g_{\mu\nu}(-\lambda-b^{-4}{B_0^2\over
2}-b^{-2}{\sigma \over \sqrt{|\gamma|}}\sum_i\delta^{(2)}(y_i))\
,\label{longitude}
\end{eqnarray}
and the transverse component is
\begin{equation}
R[g]=2b^{-1}\nabla^2b+{1\over M_6^4}(2\lambda-b^{-4}B_0^2)\
.\label{transverse}
\end{equation}

Contracting Eq.(\ref{longitude}) with $g^{\mu\nu}$ and
substituting Eq.(\ref{transverse}) into it, we can find that
\begin{equation}
b^{-2}R[\gamma]=2b^{-1}\nabla^2b+2b^{-2}(\nabla b)^2+{1\over
M_6^4}(\lambda+{3\over2}b^{-4}B_0^2+b^{-2}{2\sigma\over
\sqrt{|\gamma|}}\sum_i\delta^{(2)}(y_i))\ .\label{R2}
\end{equation}

With the help of the Gauss-Bonnet theorem,
\begin{equation}
\int d^2y\sqrt{|\gamma|} R[\gamma]=8\pi\ ,\label{gb}
\end{equation}
and from the solution (\ref{gamma}), $\mathcal{V}=4\pi
a_0^2\alpha$, we can integrate Eq.(\ref{R2}) over the transverse
coordinate. It can be found that the dependence on $\sigma$ will
just cancel out and we can get the equation of motion of radion,
\begin{equation}
2b^{-1}\nabla^2b+2b^{-2}(\nabla b)^2+{\lambda\over
M_6^4}(1-4b^{-2}+3b^{-4})=0\ .\label{radioneom}
\end{equation}

Through the conformal transformation (\ref{conformal}) and the
field redefinition (\ref{phi}), it can be checked that in the
Einstein frame, Eq.(\ref{radioneom}) just transforms to the radion
equation of motion without the brane described by the effective
potential (\ref{poten}),
\begin{equation}
\tilde{\nabla}^2\phi+{\lambda \mathcal{V}\over
M_4}(e^{-\phi/M_4}-4e^{-2\phi/M_4}+3e^{-3\phi/M_4})=0\ .\label{}
\end{equation}

Since the radion $\phi$ satisfies the same equation of motion, the
radion effective potential in the presence of branes is still
given by (\ref{poten}), i.e. the presence of branes will
contribute nothing to the radion effective potential irrespective
of the value of the brane tension $\sigma$. By the way, this
discussion also shows explicitly that after adding branes, the
model (\ref{action}) is still stable.

It is important to ask whether the above result can be generalized
to other codimension 2 brane configurations. We will argue that
there is good evidence to believe that the answer is yes. The
essential step in the above derivation is a relationship between
the sum of the brane tensions and the local curvature of the
compactification manifold following from the Gauss-Bonnet theorem.
This relationship is also directly responsible for one remarkable
feature of the solution (\ref{gamma}): if the brane tension is a
constant, the 4-dimensional geometry is independent of it
\cite{carroll-foot} (This feature has interesting implications in
addressing the cosmological constant problem. However, as recently
discussed in Ref.\cite{garriga}, if the brane tension changes
during a phase transition, the 4-dimensional geometry will fail to
remain flat. Thus the independence of the 4-dimensional geometry
on the brane tension is not a cosmologically realistic feature).
So we can see that those two features are actually intimately
related: they are both the direct results of the Gauss-Bonnet
theorem on 2-dimensional manifold with deficit angles.
Furthermore, as shown in Ref.\cite{sundrum}, applying the
2-dimensional Gauss-Bonnet theorem to more general brane
configurations, one can show that the independence of the brane
geometry on the brane tensions is a general feature of codimension
2 branes. So we think this is strong indication that the
independence of the radion effective potential on the brane
tension is actually a general feature of codimension 2 branes.

In sum, in this paper we showed that using the formula
(\ref{branepoten}) to derive the radion effective potential cannot
be always justified. The validity of the formula
(\ref{branepoten}) deserves careful examination in every specific
models.

\section*{Acknowledgement}
We would like to thank Sergei D. Odintsov and Liu Zhao for helpful
comment on the manuscript. P.W. would especially like to thank
Jeremie Vinet for clarifying discussion on the radion effective
potential. P.W. would also like to thank James M. Cline and David
Tong for helpful correspondence on this paper. X.H.M. has
benefitted a lot by helpful discussions with D.Lyth,  L.Ryder,
X.P.Wu, and X.M.Zhang. This work is supported partly by ICSC-World
Laboratory Scholarship and Doctoral Foundation of National
Education Ministry.

\end{document}